\documentclass[elsart]{article}
\input epsf.sty

\def\lsim{\lower0.6ex\vbox{\hbox{$ \buildrel{\textstyle 
<}\over{\sim}\ $}}}
\def\rsim{\lower0.6ex\vbox{\hbox{$ \buildrel{\textstyle 
>}\over{\sim}\ $}}}

\def\he3{$^{3}$He}

\def\3he{$^3$He}
\def\4he{$^4$He}
\def\6li{$^6$Li}
\def\7li{$^7$Li}
\def\7be{$^7$Be}
\def\he3{$^3$He}
\def\9be{$^9$Be}
\def\10b{$^{10}$B}
\def\11b{$^{11}$B}
\def\9b{$^9$B}
\def\12c{$^{12}$C}
\def\11c{$^{11}$C}
\def\22ne{$^{22}$Ne}
\def\n14{$^{14}$N}
\def\59ni{$^{59}$Ni}

\def\etal{{\it et al.~}}

\def\fun#1#2{\lower3.6pt\vbox{\baselineskip0pt\lineskip.9pt
  \ialign{$\mathsurround=0pt#1\hfil##\hfil$\crcr#2\crcr\sim\crcr}}}
\def\beq{\begin{equation}}
\def\eeq{\end{equation}}

\begin{document}
\begin{flushright}
astro-ph/9907171 \\
June 1999\\

\end{flushright}
\vskip 0.7in
 
\begin{center} 

{\Large{\bf Lithium-Beryllium-Boron:\\
 Origin and Evolution}}
 
\vskip 0.4in{Elisabeth Vangioni-Flam$^1$, Michel Cass\'e$^{1,2}$
and Jean Audouze$^{1,3}$}
 
\vskip 0.2in
{\it $^1${Institut d'Astrophysique de Paris, 98 bis Bd Arago 75014 Paris\\
France}}\\
\vskip 0.1in

{\it $^2${Service d'Astrophysique, DAPNIA, DSM, CEA, \\ 
Orme des Merisiers, Gif/Yvette France}}\\

{\it $^3${Palais de la D\'ecouverte, Avenue Franklin Roosevelt,  \\ 
75008 Paris, France}}\\

\vskip 1.0in
{\bf Abstract}
\end{center}

The origin and evolution of Lithium-Beryllium-Boron is a crossing point between different
 astrophysical fields : optical and gamma spectroscopy,
  non thermal nucleosynthesis, Big Bang and stellar nucleosynthesis and finally 
  galactic evolution. We describe the production and the evolution of Lithium-Beryllium-Boron
 from Big Bang up to now through the interaction of the Standard
  Galactic Cosmic Rays with the interstellar medium, supernova neutrino spallation and a low 
  energy component related to supernova explosions in galactic superbubbles.

\newpage
\noindent
\section{Introduction}

\begin{figure}[ht]
	\centering
\vspace{8.cm} 
\includegraphics{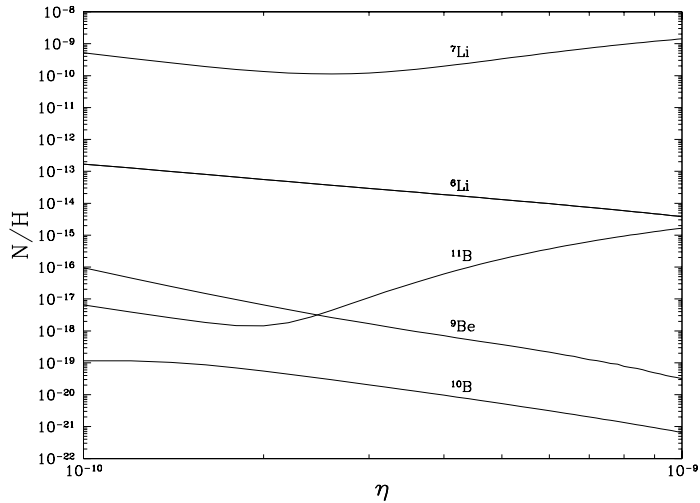}
\caption{ Big Bang Nucleosynthesis of Lithium, Beryllium and Boron vs 
    the photon over baryon ratio. N/H represents the abundance by number of
  the different nuclear species \cite{evf3}.}
	\label{fig1}
\end{figure}

Light element nucleosynthesis is an important chapter of 
nuclear astrophysics. Specifically, the rare and fragile light nuclei, 
Lithium, Beryllium and Boron (LiBeB) are not generated in the 
normal 
course of stellar nucleosynthesis (except $^{7}$Li, in the galactic disk)
 and are, in fact, 
destroyed in stellar interiors. This characteristic is reflected in the 
low 
abundance of these simple species.

A glance to the abundance curve \cite{and} suffices to capture the essence of 
the problem: a gap separates He and C. At the bottom of this 
precipice rests the trio Li-Be-B.
They are characterized by the simplicity of their nuclear structure (6 
to 11 nucleons) and their scarcity in the Solar System and in stars.

 Indeed, they are rare because they 
are fragile and, apparently a selection principle at the nuclear level 
has 
operated in nature. Due to the fact that 
nuclei with mass  5 and 8  are unstable, the Big-Bang  
nucleosynthesis (BBN)
has stopped at A = 7, and primordial thermonuclear fusion has been unable to 
proceed efficiently beyond lithium.
 Figure 1 represents the calculated 
 abundance by number of the light elements as a function of the baryon
 over photon ratio produced in the Big Bang. 

 The Big Bang production of \6li is dominated by the 
D $(\alpha, \gamma)$ \6li reaction \cite{tsof}, \cite{sch1}, \cite{sch2}, \cite{nol}.
  No direct measurement of the cross section
  of this reaction 
has been performed below 1 MeV. However, the Coulomb breakup 
technique \cite{kien} provides an indirect estimate which is
  in qualitative
 agreement with the theoretical extrapolation at low energy of Mohr \etal (1994) \cite{mohr}.
Recently, the European Collaboration between nuclear 
physicits and astrophysicits led by Marcel Arnould (NACRE : 
European Astrophysical Compilation of Reaction Rates) has 
delivered a consistent compilation of thermonuclear reaction rates 
of astrophysical interest, among them is the D  $(\alpha, \gamma)$ \6li 
reaction \cite{ang}. They conclude that the reaction rate 
based on the Mohr \etal \cite{mohr} $S$ factor is the most relevant. This 
rate is similar to that of Caughlan and Fowler (1988) \cite{caugh}, in the 
temperature range of cosmological interest. The two estimates 
agreee to within a few percents.
 Nollet \etal (1997) \cite{nol} have considered all the published 
evaluations of this reaction rate. However, most of the 
extrapolations at low temperature depart considerably from the Kiener \etal 1991
\cite{kien} estimate except that of \cite{mohr}.  
Following the recommendation of Kiener (1998, private 
communication) and \cite{ang} we have adopted the Mohr estimate.
 Note that the upper 
limit given by \cite{cecil} is much higher. This upper limit is 
indeed related to the bad sensitivity of the detector used (Kiener, 
private communication). 
   The $^{10}$B  and $^{11}$B 
abundances are calculated with updated reactions rates, including the new 
       $^{10}$B$(p,\alpha)$\7be reaction ( adopted from  \cite{raus}). Clearly,  the calculated
primordial Be and B abundances are negligibly small, compared to
 \6li. The standard BBN 
is hoplessly ineffective in generating \6li, \9be, $^{10}$B, $^{11}$B.
   \cite{sch1}, \cite{delb}).

 Thus, stars are necessary to pursue the nuclear evolution 
 bridging the gap between \4he and $^{12}$C much later, through 
nuclear fusion.

 Up to recently, the most plausible formation agents of LiBeB were thought to be
Galactic Cosmic Rays (GCRs) interacting with 
interstellar CNO. Other possible origins have been also 
identified: primordial and stellar ($^{7}$Li) and  supernova 
neutrino 
spallation (for $^{7}$Li and $^{11}$B). In contrast, \6li, \9be and $^{10}$B
  are pure spallative products. Be is very precious to astrophysics since it
 is monoisotopic,
 thus the elemental measurement is also an isotopic measurement.
 On the other hand, Li and B have two stable isotopes and an isotopic
 measurement is 
 necessary to separate 6 and 7, 10 and 11, which is very difficult,
  specifically in stars.

 \6li  presents a 
special interest since fortunatly, the \6li/$^{7}$Li ratio has been 
measured 
recently in a few halo stars, offering a new constraint on the early 
galactic evolution of light elements.
Optical measurements  of the beryllium and boron abundances 
in halo stars have been achieved by the 10 meter KECK telescope 
and 
the Hubble Space Telescope. These observations indicate a quasi 
linear 
correlation between Be and B vs Fe, at least at low metallicity (mass fraction
 of elements heavier than helium), contradictory at first sight, to a dominating GCR 
 origin of the light elements which  predicts a 
quadratic 
relationship (see appendix, section 10).
As a consequence, the theory of the origin and  evolution of 
LiBeB nuclei has to be reassessed. 
Aside GCRs, which are thought to be accelerated in the general interstellar 
medium (ISM) and which create LiBeB through the break up of interstellar CNO by their  
fast 
protons and alphas, Wolf-Rayet stars (WR) and core collapse 
supernovae (SNII) grouped in superbubbles could produce copious 
amounts of light elements via the fragmentation in flight of rapid 
carbon and oxygen nuclei (called hereafter low energy component, LEC)
  colliding with H and He in the ISM. In 
this 
case, LiBeB would be produced independently of the interstellar medium 
chemical composition and thus a primary origin is expected
 (see appendix, section 10).
These different formation  processes are discussed in the framework of a 
galactic evolutionary model. More spectroscopic observations 
(specifically of O, Fe, Li, Be, B) in halo stars are required for a better 
understanding of the relative contribution of the various mechanisms. 
Future tests on the injection and acceleration of nuclei by supernovae and 
Wolf Rayet relying on high energy astronomy will be 
invoked in 
the perspective of  X ray astronomy and the European INTEGRAL satellite.

Thus, light element research impacts several important astrophysical
problems, specifically, the origin and the evolution of cosmic
rays, galactic chemical evolution, X and gamma-ray astronomies and indirectly BBN
 through Li. 

The questions posed are: What are
the main agents of nonthermal nucleosynthesis and what are their
relative contributions at different galactic epochs? What is the
origin of the present epoch GCRs, are they accelerated out of the
average ISM or out of supernova ejecta, and what was their past
composition? Was there a population of low energy nuclei,
specifically C and O, responsible for primary LiBeB production? 

   This review is dedicated to David Schramm, our exemplary colleague 
 and friend, who has 
  largely contributed to the development of this astrophysical field.

 In section 2, we decline the story of the subject matter, in section 3
 the production ratios of the light isotopes are presented. Sections 4 and 5
 are devoted to standard Galactic Cosmic Rays, the traditional agents of 
 creation of LiBeB. Section 6 introduces and describes a new spallative 
 component made of low energy nuclei, possibly distinct from GCRs. Section 7
 briefly summarises neutrino spallation, section 8 integrates all
  these processes in the framework of a galactic evolutionary model.
 Finally, section 9 concludes and presents the prospects of LiBeB research.

\section{LiBeB story}


A general trend in nature is that complex nuclei are not proliferating: the 
abundance of the elements versus the mass number draws a 
globally 
decreasing curve. In the whole nuclear realm, LiBeB
 are exceptional since they are both simple and rare. 
Typically, in the Solar System, Li/H = 2. $10^{-9}$, B/H = 7. $10^{-10}$,
 Be/H = 2.5 $10^{-11}$ \cite{and}, whereas C/H = 3.5 $10^{-4}$
  and O/H = 8.5 $10^{-4}$.

The local isotopic ratios, measured in meteorites are known with excellent precision. 
$^{11}$B/$^{10}$B = 4 \cite{chau}, $^{7}$Li/\6li = 12.5  \cite{and}.

  Stellar nucleosynthesis, quiescent or explosive, 
forge the whole variety of nuclei from C to U but
LiBeB  nuclei are destroyed in the interior of stars, except $^{7}$Li which
 is produced in AGB and novae.
 The destruction temperatures are 2, 2.5, 3.5, 
 5.3 and 5 millions of degrees for \6li, $^{7}$Li, \9be, $^{10}$B and $^{11}$B  
  respectively.
 It is worth noting that $^{7}$Li 
 and $^{11}$B  could be produced by neutrino spallation in helium and 
 carbon shells of core
 collapse supernovae, respectively \cite{woo1}, \cite{evf2}; however,
 this mechanism is particularly uncertain depending strongly on the neutrino
 energy distribution.
 		
It is clear that another source is necessary to 
generate at least \6li, \9be, $^{10}$B and this is a non thermal mechanism, 
namely  the break up of heavier species 
(CNO, mainly) by energetic collisions, also called spallation.
  
The LiBeB story has been rich and moving. The genesis of LiBeB 
was so 
obscure to  Burbidge et al (1957) \cite{bur} that they called X the process 
leading to 
their production. Then came Hubert Reeves and his colleagues. 

 A very active and talented group of nuclear physicists, led by Ren\'e Bernas:
  Eli Gradstzjan, Robert Klapish, Marcelle Epherre, Francoise Yiou and
 Grant Raisbeck used  the mass spectroscopy techniques in order to
  determine the spallation cross sections induced by 155 MeV protons on C and
  O. H. Reeves and E. Schatzman were 
 lucky enough to collaborate with this group. The paper Bernas \etal (1967) \cite{beber}  based
 on a better knowledge of the spallation nuclear reaction cross sections has
 been determining to the understanding of the genesis of LiBeB. 

 The second happy circumstance was the encounter between Hubert Reeves and 
 Bernard Peters (one of the discoverers of the heavy Cosmic
  Rays). B. Peters draw the attention of H. Reeves on the fact 
that the LiBeB/CNO ratio
  is about $10^{-6}$ in the standard abundances (Solar System) and 0.25 
in the GCRs.
 H. Reeves proposed at once the scenario concerning the LiBeB nucleosynthesis 
 \cite{ree2}: contrary to most of the nuclear species LiBeB
 are formed by the "spallative " encounter between the energetic GCR particles
 and the interstellar medium (ISM). 
   
In a 
seminal work, Meneguzzi, Audouze and Reeves (1971) (MAR) \cite{men} gave the complete
 calculations.
 Considering the  fast p, $\alpha$ in the GCRs interacting
 with CNO in the ISM, they were able to make quantitative estimates of the 
LiBeB 
production on the basis of cross section 
measurements notably made in Orsay \cite{rai}.

 However, this study, relying on the local and present observations (LiBeB 
 and CNO abundances, cosmic ray flux and spectrum) was based on an extrapolation
 over the whole galactic lifetime assuming that all the parameters are
 constant. This result accounted fairly well for the cumulated light
 element abundances
 but obviously not for their evolution which, at that time, was unknown.
The pertinence  of the MAR proposal is illuminated by the simple and
 beautiful fact that the hierarchy of the abundances $^{11}$B $>$ $^{10}$B $>$ \9be 
is 
reflected in the cross sections \cite{rea} (see figure 4).
  This is another 
proof that nature follows the rules of nuclear physics.
\6li, \9be and $^{10}$B were nicely explained but problems were 
encountered 
with $^{7}$Li and $^{11}$B. The calculated $^{7}$Li/\6li ratio was 1.2 against 12.5 
in 
meteorites. Stellar sources of $^{7}$Li appeared necessary. The estimated 
$^{11}$B/$^{10}$B ratio was 2.5  instead of 4 in meteorites. An ad-hoc hypothesis 
drawing on unobservable low energy protons  operating through
 the \n14(p,x)$^{11}$B reaction was advocated \cite{ree1}.

New measurements of Be/H and B/H from KECK and HST, together 
with [Fe/H] \cite{rebo}, \cite{gil} \cite{dun1}
 \cite{boe1}, \cite{rya1}, \cite{dunc}, \cite{garc} in very low metallicity halo stars came to set 
strong 
constraints on the origin and evolution of  light isotopes.

\begin{figure}[ht]
	\centering
	\epsfysize=3.2truein
\epsfbox{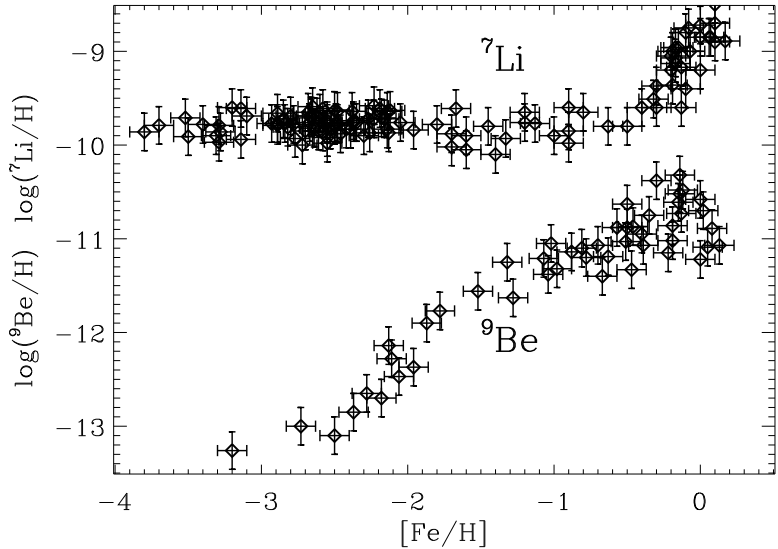}
	\caption{ Lithium (essentially $^{7}$Li) and Beryllium abundances as a function of the iron
   abundance of stars, kindly supplied by Martin Lemoine. \cite{spi, boe2, isr, boe1, gil, rya1}.}
	\label{fig2}
\end{figure}

 The evolution of BeB was suddenly uncovered  over about 10 Gyr, taking [Fe/H] 
as 
an evolutionary index.
A compilation of Be and B data is presented in figures 2 and 6. The most 
striking 
point is that log(Be/H) and log(B/H) are both quasi proportional to [Fe/H] (this notation means 
 the log of the number abundance normalized to its solar value),
 at least up 
to  [Fe/H] = -1 and that the B/Be ratio lies in the range 10 - 30
 \cite{dunc}.  Note however, two discrepant points in the
 boron diagram (figure 6) at the lowest 
 [Fe/H]. This is mainly due to the huge NLTE correction on B data \cite{kis1},
 \cite{kis2} that increases
 the departure from a straight line. It is important to take a carefull
 look to this delicate correction.  Moreover, NLTE corrections on Fe have also to be 
 considered \cite{th}.

This linearity came as a surprise since a quadratic 
relation was expected from the GCR mechanism (see section 4). It was a strong
 indication that the standard GCRs are not the main producers of LiBeB in
 the early 
Galaxy. A new mechanism of primary nature was required to
 reproduce these observations: it has been proposed that low energy CO nuclei
 produced and accelerated by massive stars (WR and SN II) fragment
  on H and He
 at rest in the ISM. This low energy component (LEC) has the advantage of
 coproducing Be and B in good agreement with the ratio observed in Pop II
 stars (figure 6, \cite{ca3}, \cite{evf2}, \cite{evf40}, \cite{ra2}, \cite{ra3}). 

 A primary origin , in this 
language, means a production rate independent of the interstellar 
metallicity (Z: mass fraction of elements heavier than helium). In this case,
 the cumulated abundance of a given light 
isotope L is approximately 
 proportional to Z. At variance, standard GCRs offer a 
secondary mechanism because it should depend both on the CNO abundance of the
 ISM at a given time and on the intensity of cosmic ray flux, itself assumed
 to be proportional to the SN II rate (see appendix, section 10).

 Indeed, the Be-Fe and B-Fe correlations taken at face value reveal
 a contradiction between standard GCR
theory and observation.
But, since oxygen is the main progenitor of BeB, the
 apparent linear relation between BeB and Fe could be misleading if O was
 not strictly proportional to Fe \cite{isr}, \cite{boe2}. Thus the pure primary origin
 of BeB in the early Galaxy 
could be questionned
 \cite{fie2}, \cite{fie3}, \cite{fie4}. However, the oxygen measurements
 themselves are confronted to many
 difficulties \cite{will}, \cite{cay3}, \cite{ful}. On the theoretical side,
 the situation is not better. The 
 [$\alpha$/Fe] vs [Fe/H] where $\alpha$ = Mg, Si, Ca, S, Ti \cite{cayr}, \cite{rya2} 
show a plateau from [Fe/H] = -4 to -1.  On nucleosynthetic
 grounds, it would be surprising that oxygen does not follow the Si and 
 Ca trends.
 Moreover, using the published nucleosynthetic yields \cite{woo2}, \cite{thie} 
it is impossible through galactic 
 evolutionary models to fit the log(O/H) vs
 [Fe/H] relation of Israelian \etal (1998) \cite{isr} and Boesgaard \etal (1999) \cite{boe2} since
 the required oxygen yields are unrealistic. Thus the subject is controversial.

Concerning lithium, a  compilation of the data is
 shown in Figure 2 \cite{lem2} and \cite{molar}. The Spite plateau extends up to [Fe/H] = -1.3.
 Beyond, Li/H is 
strongly increasing until its solar value of 2. $10^{-9}$.

 A stringent 
constraint on any theory of Li evolution is avoiding to cross the Spite's 
plateau 
below [Fe/H] = -1. Accordingly, the Li/Be production ratio should 
be 
less than about 100. 

Recent measurements of \6li have been made successfully in two halo 
stars, HD84937 and BD +26 3578 at about  [Fe/H] =  -2.3 \cite{hob1} \cite{smi1} \cite{hob2}, \cite{cay1}, \cite{smi2}, yielding
\6li/$^{7}$Li about  0.05.
The great interest of \6li,  (as shown by \cite{sfosw} ) besides of being an indicator of stellar 
destruction \cite{pin1}, \cite{cha}, \cite{cay2}, \cite{vau} is to represent a pure spallation
 product as \9be and participate to constrain the global LiBeB evolution. 
 Moreover, the \6li/\9be ratio (clearly non solar in two halo stars since
 its amounts to 20 - 80, \cite{hob4}) bears information on ancient
 non thermal nuclei.

 To summarize, we can give  six 
 observational constraints on LiBeB evolution:

                 1. Be and B proportional to Fe
 
                 2. Li/Be $<$  100 up to [Fe/H] = -1

                 3. B/Be = 10-30
 
                 4. $^{11}$B/$^{10}$B = 4 at solar birth

                 5. $^{7}$Li/\6li = 12.5 at solar birth

                 6. \6li/$^{7}$Li = 0.05  and \6li/\9be = 20 to 80 (to be
 confirmed) at [Fe/H] about -2.3

We recall that the observational O - Fe relation is central to the
 interpretation since specifically the production of Be is directly
 related to O and not to Fe.

\section{Production ratios of LiBeB isotopes}

\subsection{Cross sections}

The physics of interest for us is that of the heavy ions accelerators.
It implies a beam of fast nuclei (injected and subsequently 
accelerated), a target and an interaction between both.

The result of a nuclear collision
depends on i) the composition of the beam, ii) the relative 
velocity of the projectile and the target nuclei, and thus the energy 
spectrum of the beam  and iii)  the target composition, 
All these dependencies are quantified by the cross section.

For a monenergetic beam, the number of favourable events per 
second or reaction rate  is given by $r_{ij} = n_i n_j \sigma v$, or,
 equivalently, by 
the product  $n_i \sigma.\phi$  where  $n_i$ is the number density of the 
target and $\phi$ the flux of the projectiles.
For an energy distribution, the reaction rate is averaged on the velocity 
distribution function $\phi$(v).
In most of the astrophysical situations, the energy distribution of 
the projectile is violently non thermal. The spectra that can be 
parametrized as power laws of the form

             N(E)dE  = k$E^{-\gamma}dE$  

are customary. 
Stellar nucleosynthesis implies low energies (1 keV-100 keV) and high 
densities ($10^{2}-10^{9}$ $g.cm^{-3}$).
 By contrast, spallative nucleosynthesis 
  implies high energies (1 MeV-100 GeV) and very low densities (1-
 $10^{3}$ atoms.$cm^{-3}$). Expressed in (MeV/n) unit, the cross sections 
of direct reactions 
 (1 + 2 $->$ 3) and reverse  (2 + 1 $->$ 3) where beam and target have 
been exchanged, are identical due to the velocity conservation in energetic collisions.

In the thermonuclear context, the energy of the nuclei is below the Coulomb 
barrier, and the cross section is dominated by the Coulomb penetration factor.
  Cross sections are factorised as follows: 

         $\sigma(E) = (1/E).S(E).exp(-2\pi \eta)$

 where $\eta= Z_1 Z_2/ hv$

 The first term (square of the associated wavelength) is of geometric nature,
  the second, called astrophysical factor, is related to the internal
  structure of the 
target nucleus and the third, the Coulomb penetration factor,
  is by far the most influential.

In general, the production cross sections, and thus the production ratios
  of the 
various isotopes are very different, which explains the abundance
  disparities at 
the end of the Big Bang nucleosynthesis. In the high energy non 
thermal context, on the contrary, the penetration factor tends to 1.
  This, added to 
the nuclear similarity of light nuclei leads to spallative production rates  of
 \6li,$^{7}$Li, \9be and $^{10}$B, $^{11}$B relatively similar.

The excitation functions (variations of the cross sections as a 
function of energy/n) are presented in figure 3.

\begin{figure}[ht]
	\centering
\vspace{10.cm}
\includegraphics{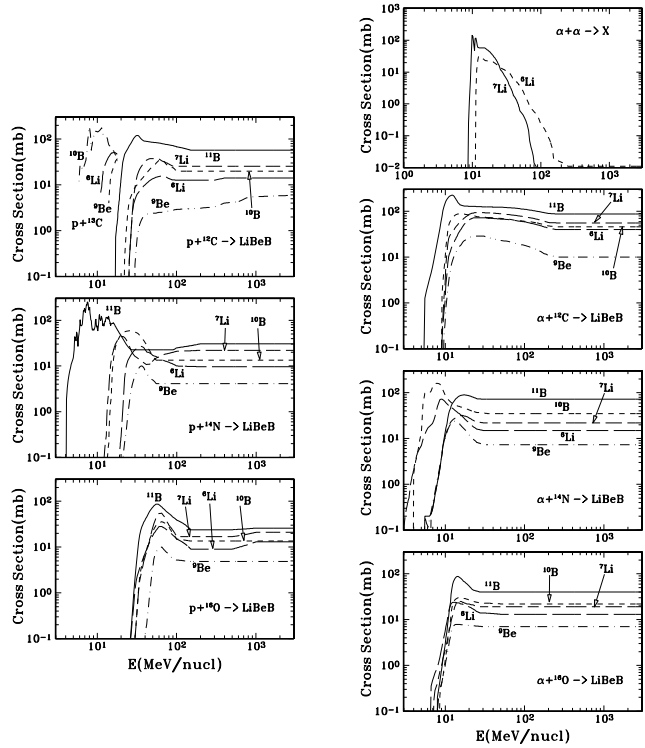}
\caption{LiBeB production cross sections as a function of the 
energy per nucleon updated by \cite{ra4}. From top to 
bottom and left to right, reactions on C, N, and O induced by protons,
  $\alpha +\alpha$ ; same reactions but with alpha particles. These graphs have been 
 kindly supplied by Reuven Ramaty.}
\label{fig3}
\end{figure}

The $\alpha+ \alpha$ reaction, leading to the synthesis of \6li and $^{7}$Li plays
  a peculiar role in astrophysics since it is the only reaction that
  implies nuclei of Big Bang origin. It is fertile even at zero metallicity
 and thus 
  it is a source of \6li and $^{7}$Li in the early Galaxy. Its excitation
  function is particular due to its low threshold and its decline at high 
energy. 
For more 
details on the nuclear physics see Reeves (1994), \cite{ree1})
 
\subsection{Input parameters}

Thus, four parameters are influential to the spallative production of light elements: 
the 
reaction cross sections, the energy spectrum of fast nuclei, the 
composition of the beam and the composition of the target. 

 Cross sections are well measured \cite{rea}
 and have been 
updated recently by \cite{ra4}.

 The adopted spectra are  generally of two kinds:

1.  GCR: N(E)dE  = k$E^{-2.7}dE$ above a few GeV/n  with a 
flatening below (e.g. \cite{lem} and \cite{mosk}). 

2. LEC : Shock wave acceleration with a cut at Eo of the form 
N(E) dE = k$E^{-1.5}$ exp(-E/Eo) dE \cite{ra2}, propagated in the ISM.

 The source composition of GCR is well determined \cite{ver}.
 It is p and $\alpha$ rich (H/O = 200, He/O = 20)
 contrary to the possible  source composition of the LEC (see section 6).
 The most obvious contributors to LEC are supernovae, 
Wolf-Rayet and mass loosing stars \cite{ca3}, \cite{ra2}, \cite{ra3}.
 It is worth noting that in the  early Galaxy,
supernovae
play a leading role since at very low metallicity stellar winds are 
unsignificant.

Table 1 shows a sample of compositions  used by different authors:
 solar system (SS) for comparison (\cite{ra2} from \cite{and}), cosmic ray source (CRS)
 (\cite{ra2} from \cite{mew}), wind of massive stars (W40) (\cite{par1} from \cite{mey}),
 composition of grain products (GR) (\cite{ra2} and \cite{lin1}), 40 Mo
 supernova at Z = $10^{-4}$ Zo from \cite{woo2}
, 35 Mo supernova of solar metallicity (\cite{ra2} from \cite{wea}).
 The two supernovae, though at different metallicities,
 (SN40 at low metallicity and SN35 at solar metallicity),  give similar yields due to the fact
 that metallicity dependent mass loss has not been taken into account in the stellar models. 
 Resulting elemental and isotopic ratios (B/Be, $^{11}$B/$^{10}$B, \6li/\9be)
 for different compositions and Eo can be found in \cite{ra2} and \cite{evf10}.

Note that N is unsignificant and that  Type II supernovae are 
O-rich whereas the winds of massive stars are C and He-rich. These 
abundance differences are important since the  highest $^{11}$B/$^{10}$B 
ratios 
are  produced by C-rich beams through$^{12}$C(p,x)$^{11}$B and the highest \6li/\9be 
 ratios are produced by He and O rich compositions.

\begin{figure}[ht]
	\centering
	\epsfysize=3.2truein 
\epsfbox{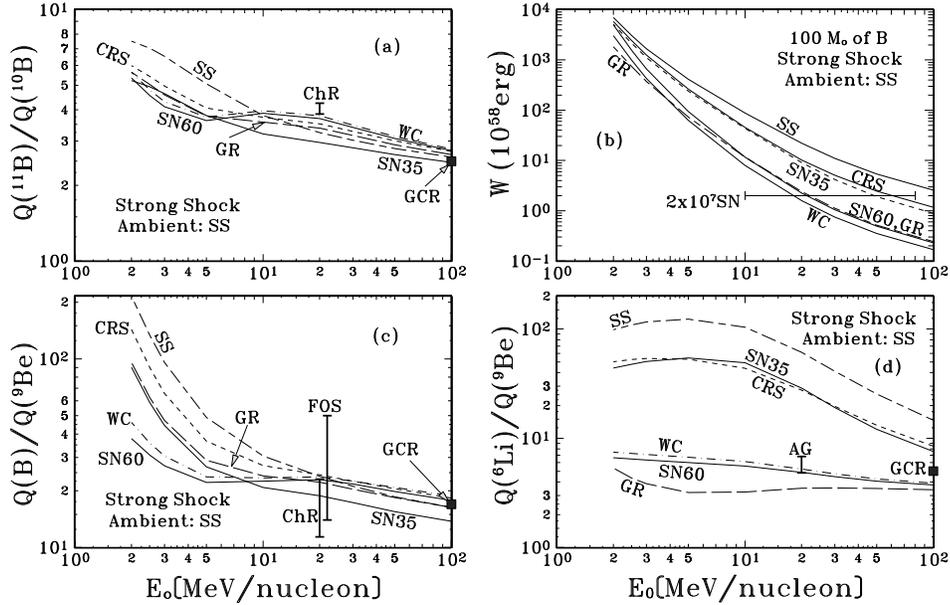}
	\caption{Production ratios of $^{11}$B/$^{10}$B (a), B/Be (c) and \6li/\9be (d)
 and energetics (b) as a function of cut-off energy Eo in the strong shock
 case, kindly supplied by Reuven Ramaty \cite{ra2}.}
   	\label{fig4}
\end{figure}

  Figure 4 shows a sample of production ratios together with the associated 
 energetics for different source compositions as a function of 
 the cut-off energy in the case of a strong shock spectrum.
 The values of the ratios at high
 Eo converge toward the GCR points. The data (ChR, FOS, AG) are from \cite{chau}
 \cite{fos} and \cite{and}. 
 The source compositions are following. CRS: cosmic ray source,
 SS: solar system, WC: Wolf-Rayet, SN35, SN60: supernovae with 35 and 60 Mo
 progenitors, GR: grains. Figure (4.b) represents the 
 energy injected under the form of fast nuclei necessary to 
 produce 100 Mo of boron (which is in order of magnitude the mass of boron
   in the whole Galaxy). The target composition is that of the present ISM taken as
 the solar one. The horizontal bar indicates the total energy released by
 $2.10^7$ SN, which corresponds roughly to the total number of SN having
  exploded since the birth of the Galaxy. 
 
\begin{table}[ht]
\centerline {\sc{\underline {Table 1 : Source Composition}}}
\vspace {0.1in}
\begin{center}
\begin{tabular}{|lcccccc|}        \hline \hline
 Element & SS	 & CRS & W40  & GR & SN40(low Z)  & SN35(Zo)  \\ \hline
H & 1200 & 220 & 80 & 2 & 37 & 27 \\
He & 120 & 22 & 25 & 0 & 8.8 & 7.6 \\
C & 0.47 & 0.87 & 1.6 & 0.3 & 0.09 & 0.08 \\
N & 0.13 & 0.04 & - & 0.03 & - & -  \\
O & 1 & 1 & 1 & 1 & 1 & 1 \\

\hline
\end{tabular}
\end{center}
\end{table}

The fourth parameter, i.e., the composition of the target (ISM) varies from
 the birth of the Galaxy up to now.
 The extensive study of 
\cite{ra2} and \cite{evf10} shows that there are only
 slight differences in the results
 when the ISM metallicity is varied between Z=0 (early galaxy) and Z=0.02 
 (at present).

\section{Standard Galactic Cosmic Rays (GCRs): acceleration and propagation }

\subsection {High energy cosmic rays}

Galactic cosmic rays represent the only sample of matter originating from
beyond the Solar System. They are constituted by bare energetic  nuclei
stripped from their electrons. Their energy density (about $1 eV cm^{-3}$ \cite{we}
  similar to that of stellar light and that of the galactic 
magnetic field), indicate that they are an important  component in the 
dynamics of the Galaxy.
A key point for our purpose is that, as said above,  GCRs
 are exceptionnally LiBeB rich 
  (LiBeB/CNO = 0.25) compared to the local (Solar  System)  matter
  ($LiBeB/CNO = 10^{-6}$).

The energy spectrum of the GCRs can be described by a power 
law above a few GeV/n (section 3.2). Below 1GeV/n, the nuclei are repelled by 
the solar wind and it is difficult to deduce from the observations 
near earth, the interstellar spectrum.

 Diffuse shock acceleration is now the leading paradigm in cosmic ray 
physics (for a review see \cite{druro}, \cite{jones}, \cite{bier}).
 In this mechanism, each passage of the 
shock front produces an energy increment. Diffusion allows 
confinement in the shock region, and after many crossings of the 
shock (both ways), the particles acquire sufficient energy 
(gyroradius) to leak out.

The energy required to sustain the GCR energy density can be supplied 
by supernovae \cite{we}. Thus SN are not necessary the sources of GCR nuclei, 
but their accceleration agents.
Recurrently models assuming that they are both the reservoir and 
  the accelerating engine have been proposed.
 The last version of this idea \cite{lin1}, though attractive
 has been criticized. This does not mean that fast nuclei 
originating from supernovae do not exist at all. If they exist they 
are different from GCR and should be called differently (see section 6).

In the simplest (linear) shock acceleration theory, the resulting 
  momentum distribution 
function f(p) is a power law of the form $p^{-3r/(r-1)}$, where r is the 
compression factor, which is a 
function of the Mach number ( $r = 4M^2 /(M^2-1$). Thus for a strong 
shock,
 (M $>>$ 1), r = 4 and f(p) is proportional to  $p^{-4}$ , which corresponds to an 
energy spectrum proportional to $E^{-2}$  at high energy. The observed 
proton spectrum  ($E^{-2.7}$) is slightly steeper due to an easier escape 
of high energy nuclei (see figure 5).
Once launched in space, these nuclei are deviated by the magnetic field 
irregularities and they lose memory of their birth place. The 
propagation is of the diffusive type. Each time they reach the 
border of the Galaxy (badly defined) they can escape or return to
the galactic disk where they suffer ionization energy losses and 
nuclear interactions.
The Galaxy, in this context is like a leaky box. The influencial 
parameter is the mean escape length, $\lambda$ (or the related confinement 
time).
At every moment on their way, the fast nuclei lose energy (by ionization 
and heating of the ISM) and can be destroyed in flight by a 
collision with other nuclei sitting in the ISM.

 The propagation equation reads

     dN(E)/dt = Q(E) - N(E)/$\tau$. d/dE [(b(E).N(E)] = 0 

(assuming a steady state). Q(E) is the source (injection) spectrum, 
b(E) is the  energy loss rate by collisions with 
ambient electrons. The loss time $\tau$ is the harmonic mean of the 
confinement time and the nuclear destruction time.
 The energy distribution is modified in the course of the 
propagation since energy losses are energy dependent.
Dividing by the mean density of the ISM one get
 
  d$\phi$(E)/dx = q(E) - $\phi$(E)/$\lambda$ + $d/dE[ \omega (E). \phi(E)]$ = 0
  
 with $\omega(E) = dE/dx (MeV/ (g.cm^{-2}))$.

The equilibrium spectrum (or interstellar spectrum) is solution of this 
equation.
It is over this spectrum and not the source spectrum that the cross sections
  have 
to be averaged to get the production rates and ratios.

\begin{figure}[ht]
	\centering
	\epsfysize=3.2truein 
\epsfbox{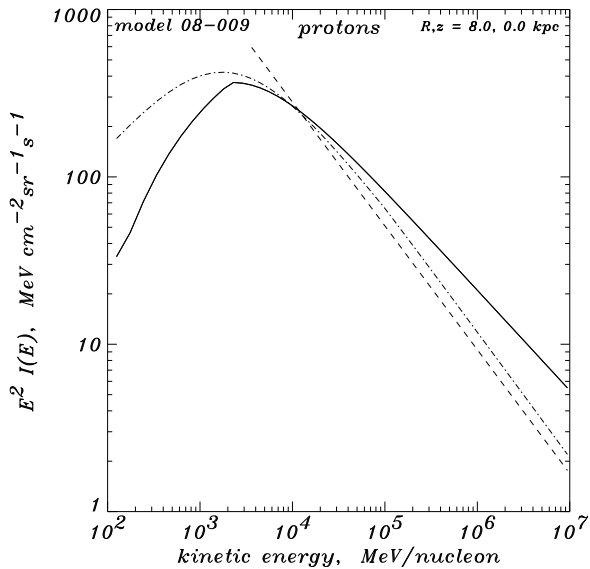}
	\caption{ Spectrum of protons calculated through a diffusion-convection
 model \cite{mosk} as a function of the kinetic energy.
 Solid line: model with no reacceleration and an injection spectrum index 2.
 Dashed line : interstellar spectrum of \cite{seo}. Dash-dotted line: 
medium spectrum \cite{mori}, kindly supplied by Andrew Strong.}
	\label{fig5}
\end{figure}

\subsection{Low energy cosmic rays and the carrot}

 At low energy (less than 100 MeV/n) the particles are thermalized 
 before leaking from the Galaxy and being destroyed by nuclear collisions.
  Thus the propagation equation simplifies considerably since the Galaxy is
 now like a closed box. In this case, the production efficiency of LiBeB 
 is maximal because fragments of nuclear collisions are quickly 
 thermalized and integrated in the interstellar gas;
 on the contrary in the high energy case most of the LiBeB 
 produced in flight escape from the Galaxy (only about 20 percent is remaining,
   \cite{men2}).

The low energy part of the problem is often overlooked, whereas the key of the 
question could lie below 100 MeV/n since the peak of the cross sections are
  about at 20 - 30 MeV/n and even lower for alpha induced reactions.
 This is due to the large uncertainties 
affecting the interstellar flux of fast nuclei in this energy range.

The propagation equation is now:

        q(E) + d/dE [$\omega$(E).$\phi$(E)] = 0

Furthermore, as we said, 
 the $^{11}$B/$^{10}$B ratio produced by GCR spallation, amounting to 2.5 
(MAR, \cite{men}) is somewhat different from the 
meteoritic value (4). MAR assumed the existence of a considerable
 excess of GCR at
low energy, (5 - 40 MeV/n). This additional flux, called carrot, if 
existing, would enhance the production of $^{11}$B via the reaction
  p + $^{14}$N $->$ $^{11}$B . 

It is worth noting that the composition of this carrot was taken as the 
GCR one and thus has to be considered as a secondary source and cannot be
 invoked to explain the linear evolution of B, indicated by the data.

The difficulty is that this component cannot be observed in 
the earth vicinity. Indeed, below about 1GeV/n, the energy distribution
 of GCR in 
interstellar space can only be determined indirectly since most particles are 
excluded from the solar cavity by the solar wind and the theory of solar 
modulation is too uncertain to demodulate the observed flux,
 especially at the lowest energies.

Limits on the low energy GCR flux (including the carrot)
 has been set by  Lemoine \etal (1998) \cite{lem} 
running a galactic evolutionary model, in order to avoid overproduction of Be.
 A 
carrot of the kind assumed by MAR and Meneguzzi and Reeves (1975) \cite{men2}
 is excluded since
 it would lead to 
strong overproduction of LiBeB at solar birth. Only the lower spectrum 
proposed in the literature was consistent with the Solar System beryllium 
abundance. Apart this global (integrated) argument there are now rather secure 
means to get informations on the low energy spectrum. 
 Strong and Moskalenko \cite{mosk}, \cite{stro1} \cite{stro2} have worked out 
a three dimensional  model of GCR propagation and confinement taking into 
account all observables (cosmic-ray composition, gamma rays, electrons, 
positrons, synchrotron radiation and antiprotons,). 
They conclude that the low energy spectrum is depressed (figure 5).
 This conclusion has an
impact on the production rate of the light elements since i) it excludes the carrot 
and ii) the decrease of the flux at low energy deduced confirms the restriction set 
by \cite{lem} on the lower end of the spectrum and it leads to a good 
fit of the local LiBeB abundances.

\section{Are Galactic Cosmic Rays a truly secondary component?}

As shown above, the CR origin impacts on the galactic evolution
 of LiBeB elements. In the light of the linear relationship between BeB and Fe,
 certain authors have asked themselves if GCRs could be primary sources, at
 the expense of modifications in the physical picture of the origin of
 the accelerated nuclei.

Indeed, since according to 
the reservoir of the accelerated nuclei (interstellar medium or
 supernova remnants 
loaded with fresh products of nucleosynthesis), the production process of LiBeB 
is secondary or primary (i.e. dependent or independent of the interstellar 
metallicity Z). A pure ISM origin would lead to a slope 2 in the correlation
 between LibeB and O (or Fe) whereas the SN origin would lead naturally to
 slope 1 (see appendix, section 10).
The debate on the origin and
evolution of the light elements has shifted i) to the
origin and very nature of the GCRs and ii) on the possible existence of
a distinct low energy component (section 6).

Concerning the status of GCRs, the question has recently 
been revived in the LiBeB Conference in Paris, December 1998 \cite{ramaty}.

Maurice Shapiro \cite{sha}, representing the traditional trend, assumed that
 the cosmic 
rays are preaccelerated by coronal mass ejection (CME) driven shocks on low mass cool 
stars and accelerated further on by passing SN shock waves. Cass\'e and Goret (1978)
\cite{ca1} first pointed out that the 
difference between the compositions of the GCR source and the solar 
photosphere might be due to selection according to the first 
ionization potential (FIP) of the elements. The FIP effect is now 
well established as the cause for the differences between the 
photospheric abundances and those of both the corona and the solar 
energetic particle in gradual events \cite{ream}. The fact that 
these energetic particles are accelerated by CME driven shocks in 
the corona is what drives the argument for the stellar origin of the 
GCR injection. However, arguments against this FIP based stellar 
origin were recently obtained from observations of ultraheavy nuclei 
in the cosmic rays \cite{wes}.

According to Lingenfelter \etal (1998) \cite{lin1}, the current epoch GCRs could come
 from each 
supernovae accelerating their own freshly produced refractory material.

According to Meyer and collaborators, the current epoch GCRs originate from 
an average ISM of solar composition as in a classical proposal since the
 composition of the external layers of stars is identical to that of that
 of the ISM where they are born. Interstellar dust plays an
 important role in determining the abundances \cite{mey2}, \cite{ell1}. In any case,
 an additional contribution 
from Wolf-Rayet circumstellar material 
enriched in freshly synthesized C and above all $^{22}$Ne is needed to explain
 the $^{22}$Ne observed in GCRs \cite{ca10}, \cite{mey1}. This WR component
 counts as a primary source, at least at present, but it is 
 quite difficult to assert its importance in the early Galaxy.

The difficulties involved in the acceleration process of the SN model 
have been emphasized by Ellison and Meyer (1999) \cite{ell2}. 
While the most significant shortcoming of this individual supernova model 
is the 
conflict with the delayed acceleration (see just below). Lingenfelter \etal (1998) \cite{lin1}
show how the averaging of the supernova nucleosynthetic 
yields over the initial mass function and supernova types, as well 
as the inclusion of the effects of refractory dust grains formed in 
the ejecta, lead to a GCR source composition in good agreement with 
the observations.
Moreover,  Meyer and Ellison (1999) \cite{mey3} stressed that the heavy s elements, and 
more specifically Sr, Zr, Ba and Ce, are not underabundant in GCR. 
They are thought to be produced in AGB stars and it is difficult to 
interpret this observation in terms of acceleration of SN material 
only.

Observations of the electron capture radioisotope $^{59}$Ni in GCR and its 
decay product $^{59}$Co \cite{bin}, have been performed with
 an instrument on the ACE 
mission. $^{59}$Ni decays by electron capture with a half life of $7.6 \times 
10^4$ years. However the decay is suppressed if the acceleration time scale is 
shorter than the lifetime because the atom is stripped as it is 
accelerated \cite{ca2}. The fact that much more 
$^{59}$Co than $^{59}$Ni is observed, suggests a delay ($\sim$10$^5$ 
years) between explosive nucleosynthesis of iron-peak nuclei (under 
the form of their radioactive progenitors) and their acceleration to 
cosmic-ray energies. This makes it unlikely that individual 
supernovae accelerate their own ejecta. This conclusion is corroborated by 
 energetic arguments.

 Obviously, the acceleration 
of average ISM matter is consistent with the implied delay between 
nucleosynthesis and acceleration. But this delay is also consistent 
with acceleration of ejecta matter in superbubbles where multiple 
supernova shocks accelerate accumulated supernova ejecta on average 
time scales at least as long as the implied delay (\cite{hig} and section 6)
 and are more energetically favourable than single supernovae \cite{pari1, pari2}.

\section{Low energy nuclei from superbubbles: a primary process}

 As soon as the Hubble and Keck observations of BeB vs Fe were released, the
 necessity of a primary component has been felt
  \cite{dun1}, \cite{evf2}, \cite{ra2}. The existence of intense fluxes of
 fast C and O nuclei in the early Galaxy appeared unescapable.
  This LEC component is, in all likelihood, physically linked to
 the superbubble scenario.

The physical conditions in superbubble cavities
would lead to fast nuclei with hard energy spectra at low energies 
up to a cutoff at an energy which is still nonrelativistic due to a combination
 of weak reflected shocks and turbulence  \cite{by3, by1,par1, par2, par3, by2}.
 These 
are the low energy nuclei which have been postulated to produce the 
bulk of the Be and B at low metallicities \cite{evf2}. 
On the other hand, as pointed out by  Higdon \etal (1998) \cite{hig, ramare}, since these giant 
superbubbles are thought to fill up a large fraction of the ISM, they are also 
the most likely site for the acceleration of the GCRs, which of 
course show no cutoff up to very high ultrarelativistic energies. 
Indeed the ingredients of a powerfull ion accelerator are 
gathered in 
superbubbles due to copious injection of matter and energy by massive stars.  
WR and SN ejections in the common pot and supersonic material in the form of 
winds and supernova ejecta able to generate recurrent shock waves. 
 Recent work by  Parizot and Drury (1999) \cite{par3} shows that SBs produce 
Be and other light elements at the less energetic cost. They could be
 the key of 
the thorny energetic problem of the production of Be in the halo. 
Superbubbles \cite{par2}, \cite{hig} thus appear as the most 
promising sites of acceleration of fresh products of nucleosynthesis, either 
as a separate low energy component or the GCRs themselves \cite{by1}, \cite{by3}.
 Indeed, both the low energy nuclei and standard 
components could be produced by superbubbles at different stages of 
their evolution.

 This idea needs further observational substantation.
Possible diagnostics of low energy particle interactions are i) non thermal 
X-rays in the 0.5 to 1 keV range due to atomic deexcitation in fast O following 
electron capture and excitation \cite{tati2}, \cite{tati3} \cite{tati5} and ii) gamma-
ray lines produced by nuclear excitation of C, O \cite{ra2}, \cite{par1}
  and $^{7}$Li + $^{7}$Be formed by the alpha + alpha reaction \cite{tati4}. 
For the time being, after the official withdrawal of the Orion gamma-ray
 results, 
\cite{blo2} and the preliminary announcement of a  detection of an 
excess between 3 and 7 MeV in the direction of Vela \cite{meu} a hint to the existence
 of a large population of MeV/n nuclei comes
 from 
the X-ray emission of the galactic ridge \cite{tati5}.
 The 
observation or non observation of C, O lines at 4.4 and 6.1 MeV and of
 the Li-Be 
feature close to 500 keV by the INTEGRAL satellite \cite{win}
 will be the  
strongest test of the superbubble hypothesis
 \cite{par1} \cite{tati4} which are at the moment the best proposed sites of acceleration of low 
energy nuclei enriched in C and O.

\section{Neutrino spallation : a primary process}

Neutrino spallation is a source of $^{7}$Li and $^{11}$B via the 
interactions of
neutrinos (predominantly $\nu_\mu$ and $\nu_\tau$) on  nuclei,
 specifically on \4he and $^{12}$C \cite{woo1}, Woosley and Weaver (1995) \cite{woo2}, hereafter WW. 
The lithium and boron yields are quite 
sensitive to the temperature of the
$\mu$ and $\tau$ neutrinos, in which there is a fair amount of
uncertainty. As a result, the overall yields of $^{7}$Li and $^{11}$B
have considerable uncertainties. 
$\nu$-process nucleosynthesis was incorporated into
a model of galactic chemical evolution (Olive \etal (1994) \cite{oli1}
 and Vangioni-Flam \etal 1996 which
 have 
included the LEC component \cite{evf2}) with the primary purpose of 
augmenting
the low value for $^{11}$B/$^{10}$B produced by standard GCR 
nucleosynthesis.
To correctly fit the observed ratio of 4, it was found that the 
yields of
WW had to be tuned down by a factor of about 2 to 5 to 
avoid 
the
overproduction of $^{11}$B. Tuning down the $\nu$-process yields ensured 
as 
well that the production of $^{7}$Li was insignificant, and did not 
affect
the Spite plateau. For a review see \cite{hart}.

Note that if taking the full yield, all galactic boron would be produced by $\nu$
 spallation. This could be a problem since \9be is not coproduced and $^{7}$Li 
 overproduced. Thus, this mechanism acts as a complement to nuclear spallative
 process at a level estimated to at most 20 percent concerning $^{11}$B, if 
one wants to fulfill the observational 
 constraints presented in section 2 \cite{evf2}.

\section{Galactic evolution of light elements}

\begin{figure}[h]
	\centering
\vspace{8.cm} 
\includegraphics{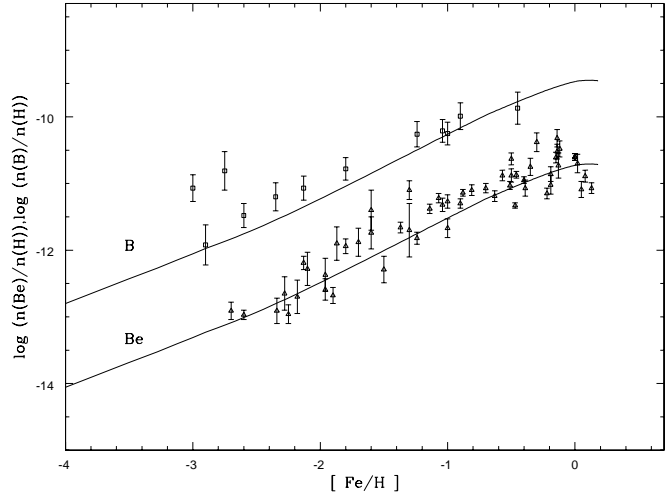}
	\caption{Beryllium and Boron evolution vs [Fe/H]. The halo evolution
 ([Fe/H] $<$ -1, is dominated by the LEC component linked to massive stars. 
 As far as B is concerned, there is room for a small contribution for $\nu$ spallation.
  \cite{evf40} \cite{evf50}}
	\label{fig6}
\end{figure}

Analyzing  all the physical parameters discussed above, two main 
 LiBeB producers emerge, the first one is the standard GCR
 in which
 fast p,$\alpha$ nuclei interact with CNO in the ISM. This process seems unable
to produce sufficient amounts of LiBeB at the level observed in the halo 
stars. However, a recent study \cite{fie2} \cite{fie3} based
 on the O - Fe relation derived by \cite{isr} and \cite{boe2} at low metallicity
 (still controversial, see section 2) try to fit the observational
 constraints with
 a pure standard GCR (secondary production) component, but has difficulties with
 the B/Be ratio and possibly with energetic constraints at very low metallicity.

The second one, LEC, invokes fragmentation of CO nuclei
 in flight by collision with H and He in the ISM.
 Massive stars  are able in principle 
 to furnish freshly synthesized C and O and accelerate them 
via the shock waves they induce in their surroundings. This mechanism is probably related to superbubbles.

Finally, neutrino spallation is helpful to increase the $^{11}$B/$^{10}$B ratio 
up to the value 
observed in meteorites. 
It is also a primary process since it implies the break up of C 
within 
supernovae and not in the ISM. However it cannot be the unique 
mechanism to produce light elements since it does not make \9be. 

These different mechanisms are included in a galactic 
evolutionary model to follow the whole
 evolution of each isotope. 

In brief, the characteristics of the standard  galactic evolution
 model used \cite{au, evf1, evf2, evf40, 
 fie2, fie3} are the following:

 No instant recycling approximation, in other terms account is taken of the time 
delay  of matter ejection by low mass stars.
 The cosmic ray flux and that of the LEC are taken proportional to the supernova rate
(itself proportional to the star formation rate). 
The GCR production rate is time dependent through the growth of CNO in the ISM which
 is 
followed by the model (equiped with the relevant yields).
The  mass distribution of stars  at  birth,  or  initial mass function is
proportional to $M^{-2.7}$ from  0.4 to 100 Mo as usual.
  The star formation rate (SFR) is taken proportional to the gaz mass fraction all 
along the galactic life.
The stellar lifetimes are taken from \cite{mey}.
¥  The composition of the ejecta is that calculated by WW.
  The model, of course, takes into account the destruction of LiBeB in stars.
¥ The  SNI rate is taken  constant, with the value observed today.
This simple model is sufficient to map the evolution of each Light isotope in the 
abundance-abundance diagram (L-O and L-Fe plots).

The theoretical evolutionary curve of Be is normalized on the 
solar abundance of this element.

 Concerning beryllium and boron, in this context,   
the main results are the following: the quasi-linearity (Be-B vs Fe) is
easily  reproduced  thanks to the action of the LEC (fig 6, \cite{evf40}).
Standard GCRs contribute no more than about 30 per cent to Solar System values. 
The B/Be ratio is in the range 10-30 as observed. The value 30 leaves enough 
room for neutrino spallation to reach $^{11}$B/$^{10}$B  = 4 at solar birth,
 but this contribution is 
 marginal \cite{evf2}.

The \6li/H ratio can also  be explained in the 
framework 
of the same model \cite{evf3}, this
without piercing the Spite plateau (figure 7). In this figure, showing
 the evolution of
 \6li/H vs [Fe/H], it can be seen that GCR is overwhelmed by LEC.
 The decrease of the  \6li/\9be
 ratio could be  explained in terms of the variation of the composition of 
superbubbles in the course of the galactic evolution, being O rich at 
start 
due to SNII and becoming more and more C rich due to the 
increasing contribution of mass loosing stars.

  Fields and Olive \cite{fie2, fie3} using the new O-Fe correlation and considering only
 the GCR component have also reproduced both solar and popII 
 $^{6}$Li in a quite natural way.  The proportions between the different
 production processes could be modified if the new O-Fe 
relation is verified but even in this case, a primary 
component oparating in the
  very early galactic phase cannot be excluded.

\begin{figure}[h]
	\centering
\vspace{8.cm} 
\includegraphics{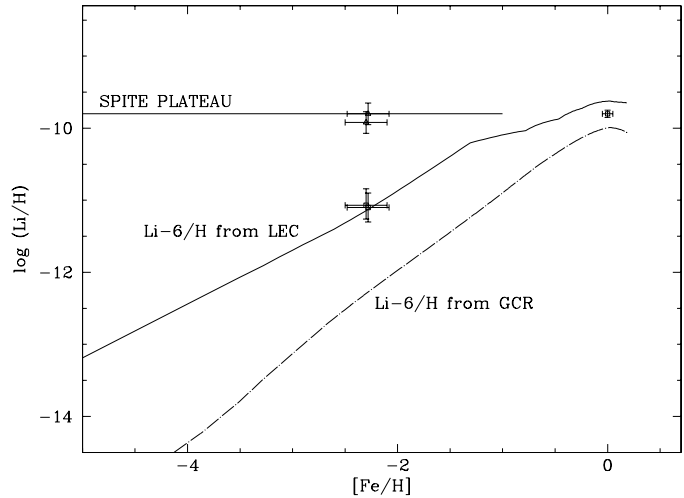}
	\caption{\6li and $^{7}$Li evolution vs [Fe/H]. Horizontal full line:
 Li abundance, essentially $^{7}$Li. Full oblique line: LEC contribution to \6li. 
 Data points concern the two observed stars, (total Li and \6li respectively).
 Dash-dotted line: GCR contribution to the \6li abundance taking the 
classical flat O/Fe relation.\cite{evf3}.}
	\label{fig7}
\end{figure}

 In both scenarios,
 the evolutionary curve of \6li crosses the
 halo observations
  meaning that \6li is almost essentially intact in the envelope
 of stars in
 which it is measured. $^{7}$Li in turn, more tightly bound than \6li,
 is even less destroyed,
 thus the mean value of the Spite plateau reflects nicely the Big Bang $^{7}$Li
 abundance. This reinforce the use of $^{7}$Li as a cosmological baryometer as shown also recently
 by the observational and theoretical analyses of \cite{molarette} \cite{rnb} \cite{rbofn}.

\section{Conclusion and perspectives}

The confrontation of theory with observations, and the crossing of
the boundaries of experiment and theory, has permitted
theorists, observers and experimentalists to clarify
fundamental questions on the origin and evolution of light elements.
 Different evolutionary scenarios
concerning LiBeB production by accelerated particles in the Galaxy,
 remain viable.

1. The delay ($\sim$$10^5$ years) between nucleosynthesis and
acceleration, implied by the $^{59}$Co and $^{59}$Ni cosmic-ray
data, is consistent with both acceleration of average ISM matter
or ejecta matter in superbubbles.

2. A low energy component distinct from standard GCRs appears necessary,
 specifically in the early Galaxy,
 but needs to be more observationally rooted.

3. Future X-ray and gamma-ray line observations should help to
clarify the very existence of the low-energy nuclei which are able
to generate LiBeB in a primary mode, and if these low energy cosmic
rays indeed exist, determine their energy spectrum. In spite of the
problems encountered in the observation and analysis of the Orion
region, the Vela region should be a prime target for the INTEGRAL
mission.

4. The new [O/Fe] vs. [Fe/H] observations may make it
necessary to refine the model in which GCR nucleosynthesis, taken
as a secondary process, is combined with nucleosynthesis by low
energy nuclei originating from superbubbles, which is a primary
process. The contribution of each component will have to be
adequately rescaled to account for the enhanced contribution of
the secondary process; the contribution of neutrino spallation would
also depend on this rescaling, but it will stay, in any case, marginal.

5. New abundance measurements of \6li close to [Fe/H] = $-2.3$, 
integrated in the framework of galactic evolutionary models, 
indicate that this nucleus is not significantly depleted in the 
observed stars and therefore Li (mostly $^7$Li which is less fragile 
than $^6$Li) is essentially intact in halo stars (Spite plateau). 
This makes Li an excellent indicator of the baryonic density of the 
Universe. The \6li Galactic evolution can be made consistent with 
the other light nuclei.

Thus, the simplistic vision of the origin and evolution of LiBeB has
developed into a complex array of possibilities due to the wealth of
observational discoveries. Nevertheless, the originally proposed
fundamental mechanism remains preserved, namely the nuclear spallative 
 process which is the only one to coproduce the light isotopes in the 
 right proportion.

The needs for the future are the following:
 
 On the theoretical side it would be necessary:

 i) to check 
NLTE corrections on B abundances (and perhaps Fe) at very low Z.

 ii) to develop and refine SN II models, especially at very low Z and 
high mass.

 On the observational side, it would be desirable to get 
measurements 
of \6li, $^{7}$Li, \9be, $^{10}$B, $^{11}$B, O, Fe in
 the same halo stars and to get \6li/$^{7}$Li and
  $^{11}$B/$^{10}$B 
ratios in stars of both populations (popI and pop II). 

Finally, the 
observation of C,O and Li-Be gamma-ray lines are important 
objectives of the INTEGRAL satellite following the
 road open by the COMPTON GRO satellite.

\section{Appendix}

A clear distinction should be made between a primary production mechanism and
 a secondary one, in the sense of the galactic evolution. Indeed, in the
 plausible hypothesis that the GCR flux $\phi$ is proportional to the
 supernova rate dN(SN)/dt, the rate of increase of L/H, L standing for 
 beryllium or boron is:

    $d(L/H)/dt = z(t) \langle \sigma \phi(t) \rangle$

 where z(t) is the evolving CNO fraction by number and $\langle \sigma. \phi(t) \rangle$
 the energy averaged of the production cross section times the flux. Since 
 z is proportional to N(SN), the integrated number of SN up to time t, and
 $\phi$ is proportional to dN(SN)/dt, we have, assuming a constant spectral shape,

  $d(L/H)/dt \alpha N(SN) dN(SN)/dt$ or $L/H \alpha z^2$

On the contrary, freshly synthesized C and O nuclei injected/accelerated in SN
 ejecta or superbubbles would lead to a primary production of LiBeB,
 independent of the interstellar medium metallicity, by fragmentation of these
 species on the surrounding H and He nuclei. So the production rate is
  independent of metallicity z, and the cumulated abundances L/H are 
proportional to z.

For example, to illustrate our purpose, we calculate
the production rate of Be by protons fragmenting interstellar CNO by the GCR process:

		 $dN(Be)/dt = N(CNO)\sigma \phi$

Dividing by N(H), the hydrogen number density, one gets the
 rate of increase  of 
the Be/H ratio as a function of z defined as N(CNO)/N(H): 

 d/dt(Be/H) = $z(t).\sigma \phi(t)$

The present rate of increase of Be/H, calculated from realistic
 numerical values,
 $N(CNO)/N(H) = 10^{-3}, \sigma = 5 mb, \phi =  10 cm^{-2}$, 
for E $>$ 30 MeV is about $10^{-28} s^{-1}$,
  which integrated over the galactic lifetime
($10^{10}$ yrs) leads to Be/H about $10^{-11}$ in qualitative agreement
 with the Solar System value, which is encouraging.
 However, this rough calculation does not take 
into account the temporal variation of the parameters.

\vskip .2in

\noindent {\bf Acknowledgments}

We thank warmly Martin Lemoine, Keith Olive, Reuven Ramaty and Andrew Strong
 for their help in the elaboration of this article. We are also grateful to
 them for their precious  scientific collaboration. We are undebted to
  Roger Cayrel, Alain Coc, Brian Fields,  Jurgen Kiener, Roland Lehoucq, Yvette Oberto, 
 Monique Spite, Francois Spite,
 Vincent Tatischeff  and Fr\'ed\'eric Th\'evenin for their active participation
 to the LiBeB saga.
   Finally many fruitful discussions have been conducted with Hans Bloemen, 
Andrei Bykov, Douglas Duncan, 
Donald Ellison, Nicolas Grevesse, Andr\'e Maeder, Jean Paul Meyer,
 Georges Meynet, Paolo Molaro,
 Etienne Parizot, Francesca Primas, Stan Woosley, thanks to them.

To close this review, we would like to pay a last homage to Dave Schramm; his vision 
 to bring together astrophysics, nuclear physics, particle physics and cosmology
 has had a true impact on our understanding of the Universe.

This work was supported in part by PICS no 319 CNRS.


\begin{thebibliography}{}



\bibitem{and} E. Anders, and N. Grevesse, Geochim. Acta 35 (1989) 197

\bibitem{au} J. Audouze, and B. Tinsley, Ap.J. 192 (1974) 487

\bibitem{blo2} H. Bloemen,  \etal, (1999), in The Extreme Universe, 3rd
INTEGRAL Workshop, Taormina, Sicilia, in press

\bibitem{boe2} A.M. Boesgaard, J.R. King, C.P. Deliyannis, and S.S.
Vogt, AJ 117 (1999) 492

\bibitem{beber} R. Bernas \etal, Ann. of Phys. 44 (1967) 426 

\bibitem{bur} E.M. Burbidge, G.B. Burbidge, W.A. Fowler, and F.Hoyle,
 Rev. Mod. Phys. 29 (1957)  547


\bibitem{ang} C. Angulo, \etal, At. Nucl. Data Tables (1999) in press

\bibitem{caugh} G.R. Caughlan, and W.A. Fowler, At.Nucl.Data Tables 
40 (1988) 283

\bibitem{cecil} F.E. Cecil, J. Yan, and C.S. Galovich, Phys.Rev.C. 
53 (1996) 1967

\bibitem{cha} B. Chaboyer, (1999)  astroph/9803106

\bibitem{delb} P. Delbourgo-Salvador and E. Vangioni-Flam, (1993) 
in " Origin and
 Evolution of Elements", Edts Prantzos \etal, Cambridge 
University Press, p. 52


\bibitem{kien} J. Kiener,  \etal,  Phys.Rev.C. 44 (1991)  2195


\bibitem{mohr} P. Mohr, \etal, Phys.Rev.C. 50 (1994) 1543

\bibitem{nol} K.M. Nollet, M. Lemoine, and D.N. Schramm, 
Phys. Rev. C56 (1998) 1144

\bibitem{raus} T. Rauscher, and G. Raimann, (1997) nucl-th/9602029

\bibitem{sch1} D.N. Schramm,(1993) in "Origin and Evolution of the
 Elements", Edts
 Prantzos \etal , Cambridge University Press, p. 112

\bibitem{sch2} D.N. Schramm,  (1994) in "Light Element Abundances", Edts P. Crane, ESO
Astrophysics Symposia, Springer, p. 50


\bibitem{tsof} D. Thomas, D.N. Schramm, K.A. Olive, and B.D. Fields, Ap.J. 406 (1993)  569 


\bibitem {bin} R. Binns, (1999) in LiBeB Cosmic Rays and Gamma-Ray Line Astronomy, ASP Conference Series
 Vol.171 eds R. Ramaty, E. Vangioni-Flam, M. Cass\'e and K. Olive, to appear

\bibitem{by1} A.M. Bykov, Space Sci. Rev 74 (1995) 397

\bibitem{by2} A.M. Bykov, and G.D. Fleishman, 1992, MNRAS 15 (1992) 269

\bibitem{by3} A. M. Bykov, (1999) in LiBeB Cosmic Rays and Gamma-Ray Line Astronomy,
 ASP Conference Series
 Vol.171 eds R. Ramaty, E. Vangioni-Flam, M. Cass\'e and K. Olive, to appear 

\bibitem{cay3} R. Cayrel, (1999) in LiBeB Cosmic Rays and Gamma-Ray Line Astronomy,
 ASP Conference Series
 Vol.171 eds R. Ramaty, E. Vangioni-Flam, M. Cass\'e and K. Olive, to appear
 
\bibitem{ell2} D.C. Ellison and J.P. Meyer, (1999) in LiBeB Cosmic Rays and
 Gamma-Ray Line Astronomy,
 ASP Conference Series
 Vol.171 eds R. Ramaty, E. Vangioni-Flam, M. Cass\'e and K. Olive,, to appear

\bibitem{fie4} B.D. Fields and K.A. Olive, (1999) in LiBeB Cosmic
 Rays and Gamma-Ray Line Astronomy,
 ASP Conference Series
 Vol.171 eds R. Ramaty, E. Vangioni-Flam, M. Cass\'e and K. Olive, to appear

\bibitem{hart} D. Hartmann, (1999) in LiBeB Cosmic Rays and Gamma-Ray Line Astronomy,
 ASP Conference Series
 Vol.171 eds R. Ramaty, E. Vangioni-Flam, M. Cass\'e and K. Olive, to appear 

\bibitem{mey3} J.P. Meyer and D.C. Ellison, (1999) in LiBeB Cosmic Rays and Gamma-Ray Line Astronomy,
 ASP Conference Series
 Vol.171 eds R. Ramaty, E. Vangioni-Flam, M. Cass\'e and K. Olive, to appear

\bibitem{hob4} L.M. Hobbs, (1999) in LiBeB Cosmic Rays and Gamma-Ray Line Astronomy,
 ASP Conference Series
 Vol.171 eds R. Ramaty, E. Vangioni-Flam, M. Cass\'e and K. Olive, to appear

\bibitem{molarette} P. Molaro, (1999) in LiBeB Cosmic Rays and Gamma-Ray Line Astronomy,
 ASP Conference Series
 Vol.171 eds R. Ramaty, E. Vangioni-Flam, M. Cass\'e and K. Olive, to appear 
 
\bibitem{sha} M.M. Shapiro, (1999) in LiBeB Cosmic Rays and Gamma-Ray Line Astronomy,
 ASP Conference Series
 Vol.171 eds R. Ramaty, E. Vangioni-Flam, M. Cass\'e and K. Olive, to appear


\bibitem{ramare} R. Ramaty and R.E. Lingenfelter (1999) in LiBeB Cosmic Rays and Gamma-Ray Line Astronomy,
 ASP Conference Series
 Vol.171 eds R. Ramaty, E. Vangioni-Flam, M. Cass\'e and K. Olive, to appear

\bibitem{kis2} D. Kiselman, (1999) in LiBeB Cosmic Rays and Gamma-Ray Line Astronomy,
 ASP Conference Series
 Vol.171 eds R. Ramaty, E. Vangioni-Flam, M. Cass\'e and K. Olive, to appear

\bibitem{stro2} A.W. Strong and I.V. Moskalenko, (1999)  in LiBeB Cosmic Rays 
and Gamma-Ray Line Astronomy,
 ASP Conference Series
 Vol.171 eds R. Ramaty, E. Vangioni-Flam, M. Cass\'e and K. Olive, to appear 


\bibitem{ramaty} R. Ramaty, E. Vangioni-Flam, M. Cass\'e and K. Olive (1999):
 Editors of the Conference on
 LiBeB Cosmic Rays and Gamma-Ray Line Astronomy, Paris December 1998, ASP Conference Series
 Vol.171, to appear

\bibitem{hob3} L.M. Hobbs and V.A. Thorburn, Ap.J. (1999) to be published

\bibitem{ca1} M. Cass\'e, and Ph. Goret, Ap.J. 221 (1978) 703

\bibitem{ca10} M. Cass\'e and J. Paul, Ap.J. 258 (1982) 860

\bibitem{ca2} M. Cass\'e, and A. Soutoul, Ap.J. 200 (1975) L75

\bibitem{ca3} M. Cass\'e, R. Lehoucq, and E. Vangioni-Flam,
Nature 373 (1995) 38

\bibitem{cay1} R. Cayrel, M. Spite, F. Spite, E. Vangioni-Flam, M. 
 Cass\'e, and J. Audouze, A.A. 343 (1999) 923 astrop-ph/9901205


\bibitem{cay2} R. Cayrel, Y. Lebreton, and P. Morel,  (1999) in Galaxy
Evolution: connecting the distant Universe with the local fossil
record, Eds. M. Spite \& F. Crifo, (Dordrecht: Kluwer), in press, 
astro-ph/9902068

\bibitem{evf50} E. Vangioni-Flam and M. Cass\'e, (1999) in Galaxy
Evolution: connecting the distant Universe with the local fossil
record, Eds. M. Spite \& F. Crifo, (Dordrecht: Kluwer), in press, astro-ph/9902073

\bibitem{chau} M. Chaussidon and F. Robert, Nature 374 (1995) 337

\bibitem{dun1} D.K. Duncan, D.L. Lambert, and M. Lemke,  Ap.J. 401 (1992) 584

\bibitem{bier} P.L. Biermann, A.A. 271 (1993) 649

\bibitem{ell1} D.C. Ellison, L.O'C. Drury, and J.P. Meyer, Ap.J. 487 (1997) 197

\bibitem{pari1} E. Parizot and L.O'C. Drury, A.A. 346 (1999) 329

\bibitem{pari2} E. Parizot and L.O'C. Drury, A.A. 346 (1999) 686

\bibitem{druro} L.O'C. Drury, Rep. Prog. Phys. 46 (1983) 973

\bibitem{jones} F.C. Jones and D.C. Ellison, Sp. Sci. Rev. 58 (1991) 259

\bibitem{fie2} B.D. Fields, and K.A. Olive,  New Astronomy (1999) in press, astro-ph/9811183

\bibitem{fie3} B.D. Fields, and K.A. Olive, Ap.J. 516 (1999), astro-ph/9809277

\bibitem{fos} B.D. Fields, K.A. Olive, and D.N Schramm, Ap.J. 435 (1994) 185

\bibitem{ful} J.P. Fulbright and R.P. Kraft, AJ (1999) to be published

\bibitem{hig} B. Higdon, R.E. Lingenfelter, and R. Ramaty, Ap.J. 509 (1998) L33

\bibitem{hob1} L.M. Hobbs, and V.A. Thorburn, Ap.J. 428 (1994)  L25

\bibitem{hob2} L.M. Hobbs, and V.A. Thorburn, Ap.J. 491 (1997)  772

\bibitem{isr} G. Israelian, R.J. Garcia Lopez, and R. Rebolo, Ap.J. 507 (1998) 805


\bibitem{lem} M. Lemoine, E. Vangioni-Flam, and M. Cass\'e, Ap.J. 499 (1998) 735

\bibitem{lin1} R.E. Lingenfelter, R. Ramaty, and B. Kozlovsky, Ap.J. 500 (1998) L153

\bibitem{men2} M. Meneguzzi and H. Reeves, A.A. 40 (1975) 91 

\bibitem{men} M. Meneguzzi, J. Audouze, and H. Reeves, A.A. 15 (1971) 337

\bibitem{mey1} J.P. Meyer, Ap.J.Supp. 57 (1985) 173

\bibitem{mey2} J.P. Meyer, L.O'C.  Drury, and D.C. Ellison, Ap.J. 487 (1997) 182

\bibitem{mori} M. Mori, Ap.J. 478 (1997) 225

\bibitem{mosk} I.V. Moskalenko and A.W. Strong, Ap.J. 493 (1998) 694

\bibitem{oli1} K.A. Olive, N. Prantzos, S. Scully, and E. Vangioni-Flam, Ap.J. 424 (1994) 666

\bibitem{par1} E. Parizot, M. Cass\'e, and E. Vangioni-Flam, A.A. 328 (1997) 107

\bibitem{par2} E. Parizot, A.A. 331 (1998)  726

\bibitem{par3} E. Parizot and L.O'C. Drury, A.A. (1999) to be published

\bibitem{rai} G. Raisbeck, and F. Yiou, F., Phys. Rev. A 4 (1971) 1848

\bibitem{ra2} R. Ramaty, B. Kozlovsky, and R.E. Lingenfelter, Ap.J. 456 (1996) 525

\bibitem{ra3} R. Ramaty, B. Kozlovsky, and R.E. Lingenfelter, Physics Today 51 No. 4 (1998) 30

\bibitem{ra4} R. Ramaty, B. Kozlovsky, R.E. Lingenfelter, and H.
Reeves, Ap.J. 488 (1997) 730

\bibitem{rea} S. Read, and V. Viola, Atomic Data Nucl. Data Tables 31 (1984) 359

\bibitem{ream} D.V. Reames, Adv. Space Res. 15 No. 7 (1995) 41

\bibitem{ree1} H. Reeves, Rev. Mod. Phys. 66 (1994) 193

\bibitem{ree2} H. Reeves, W.A. Fowler, and F. Hoyle, Nature 226 (1970) 727

\bibitem{seo} E.S. Seo \etal, Ap.J. 378 (1991) 763

\bibitem{smi1} V.V. Smith, D.L. Lambert, and P.E. Nissen, Ap.J. 408 (1993) 262

\bibitem{smi2} V.V. Smith, D.L. Lambert, and P.E. Nissen, Ap.J. 506 (1998) 405

\bibitem{stro1} A.W. Strong, and I.V. Moskalenko, Ap.J. 509 (1998) 212

\bibitem{tati1} V. Tatischeff, R. Ramaty, and B. Kozlovsky, Ap.J. 504 (1998) 874

\bibitem{tati2} V. Tatischeff, and R. Ramaty, Ap.J. 511 (1999) 204

\bibitem{tati5} V. Tatischeff, R. Ramaty and A. Valiania, (1999) Astro-ph/9903326

\bibitem{tati3} V. Tatischeff, and R. Ramaty, (1999) in The Extreme
 Universe, 3rd
INTEGRAL Workshop, Taormina, Sicilia, in press

\bibitem{tati4} V. Tatischeff \etal, (1999) in preparation

\bibitem{th} F. Th\'evenin, and T. Idiart, Ap.J. (1999) in press

\bibitem{meu} R.P. Van der Meulen, \etal (1999) in The Extreme Universe, 3rd
INTEGRAL Workshop, Taormina, Sicily, in press

\bibitem{evf1} E. Vangioni-Flam, M. Cass\'e, J. Audouze, and Y. Oberto, Ap.J. 364 (1990) 568

\bibitem{evf2} E. Vangioni-Flam, M. Cass\'e, B.D. Fields and  K.A. Olive, Ap.J. 468 (1996) 199

\bibitem{evf3} E. Vangioni-Flam, M. Cass\'e, R. Cayrel, J. Audouze, M. Spite, F. Spite,
  New Astronomy, (1999) in press astro-ph9811327

\bibitem{evf40} E. Vangioni-Flam, R. Ramaty, K.A. Olive and M. Cass\'e, A.A. 337 (1998) 714

\bibitem{we} W.R. Webber, Sp. Sci. Rev. 81 (1997) 407

\bibitem{wes} A.J. Westphal, P.B. Price, B.A. Weaver, and V.G. Afanasiev, Nature 396 (1998) 50

\bibitem{woo1} S.E. Woosley, D.H. Hartmannn, R.D. Hoffman, and W.C.
Haxton, Ap.J. 356 (1990) 272

\bibitem{woo2} S.E. Woosley, and T.A. Weaver, Ap.J.Supp. 101 (1995) 181

\bibitem{tsof} D. Thomas, D. Schramm, K.A. Olive, and B. Fields,
 Ap.J. 406 (1993) 569.

\bibitem{spi} F. Spite, and M. Spite,  A.A.  115 (1982) 357; 
M. Spite, J.P. Maillard, and F. Spite,  A.A.141 (1984) 56; 
F. Spite, and M. Spite,  A.A. 163 (1986) 140;
L.M. Hobbs, and D.K. Duncan,  Ap.J. 317 (1987) 796;
R. Rebolo, P. Molaro, J.E. and Beckman, A.A. 192 (1988) 192;
M. Spite, F. Spite, R.C. Peterson, and F.H. Chaffee Jr., 
 A.A. 172 (1987) L9; R. Rebolo, J.E. Beckman, and P. Molaro,
A.A. 172 (1987) L17; L.M. Hobbs, and C. Pilachowski, 
Ap.J. 326 (1988) L23;
L.M. Hobbs, and J.A. Thorburn, Ap.J. 375 (1991) 116;
J.A. Thorburn, Ap.J. 399 (1992) L83;
C.A. Pilachowski, C. Sneden,and J. Booth, Ap.J. 407
(1993) 699; 
 L. Hobbs, and J. Thorburn, Ap.J.  428 (1994) L25;
 J.A. Thorburn, and T.C. Beers, Ap.J. 404 (1993) L13;
F. Spite, and M. Spite, A.A.  279 (1993) L9.
J.E. Norris, S.G. Ryan, and G.S. Stringfellow, Ap.J. 423 (1994) 386.

\bibitem{rnb} S. Ryan, J. Norris, and T. Beers, (1999) astro-ph/9903059.

\bibitem{pin1} M.H. Pinsonneault, T.P. Walker, G. Steigman, and V.K.
Naranyanan, astro-ph/9803073.

\bibitem{sfosw}G. Steigman, B. Fields, K.A. Olive, D.N. Schramm, and 
T.P.  Walker, Ap.J. 415 (1993) L35

\bibitem{smi1}V.V. Smith, D.L. Lambert, and P.E. Nissen,  Ap.J. 408
(1992) 262; Ap.J. 506
(1998) 405; L. Hobbs, and J. Thorburn, 
Ap.J. 428 (1994) L25;  Ap.J. 491 (1997) 772;  R. Cayrel,
M. Spite, F. Spite, E. Vangioni-Flam, M. Cass\'e,
and J. Audouze,  A.A. 343 (1999) 923.

\bibitem{rbofn} S. Ryan, T. Beers, K.A. Olive, B.D. Fields, and J. Norris, (1999)
astro-ph/9905211.

\bibitem{boe1} A.M. Boesgaard, and J.R. King, AJ 106 (1993) 2309

\bibitem{cayr} R. Cayrel, A.A.Rev 7 (1996)  217

\bibitem{dunc} D.K. Duncan et al, Ap.J. 488 (1997) 338

\bibitem{ver} M.A. Du Vernois, (1996) in "Cosmic Abundances", Edts ASP Conferences series, 
 Vol 99, p.385

\bibitem{garc} R.J. Garc\'ia-L\'opez, \etal, Ap.J. 500 (1998) 241

\bibitem{gil} G. Gilmore, B. Gustafsson, B. Edvardsson, and P.E. Nissen, Nature 357 (1992) 379

\bibitem{isr} G. Israelian, R.F. Garc\'ia-L\'opez, and R. Rebolo, Ap.J. 507 (1998) 357

\bibitem{kis1} D. Kiselman, and M. Carlsson, A.A. 311 (1996) 681

\bibitem{lem2} M. Lemoine, D.N. Schramm, J.W. Truran, and C.J. Copi, Ap.J. 478 (1997) 554

\bibitem{will} A. Mac Williams, Ann.Rev.Astron.Astrophys. 55 (1997) 503

\bibitem{mew}  R. Mewaldt, Rev.Geophys.Space Phys. 21 (1983) 295

\bibitem{mey} G. Meynet, \etal, A.A.S 103 (1994) 97 

\bibitem{molar} P. Molaro, P. Bonifacio, F. Castelli, and L. Pasquini, A.A. 319 (1997) 593

\bibitem{rebo} R. Rebolo, P. Molaro, J.E. Beckman, A.A. 192 (1988) 192

\bibitem{rya1} S.G. Ryan, I. Norris, M. Bessel, and C. Deliyannis, Ap.J. 388 (1994) 184
\bibitem{rya2} S.G. Ryan, J.E. Norris, and T.C. Beers, Ap.J. 471 (1996) 254

\bibitem{thie} F.K. Thielemann, K. Nomoto, and M. Hashimoto, Ap.J. 460 (1996) 108 

\bibitem{evf10} E. Vangioni-Flam, M. Cass\'e, and R. Ramaty, (1997) in Proceedings
 of the 2nd INTEGRAL Workshop, "The Transparent Universe", Saint Malo, 
  France, ESA, SP.382 p. 123

\bibitem{vau} S. Vauclair, and C. Charbonnel, A.A. 502 (1998)  372

\bibitem{wea} T.A. Weaver, and S.E. Woosley, Phys.Rept 227 (1993) 65

\bibitem{win} C. Winkler, (1997) in "The Transparent Universe", 2nd INTEGRAL
 Workshop, Saint Malo, Edts, ESA, SP-382, p. 573


\end{thebibliography}
\end{document}